\documentclass[12pt]{article}
\begin{document}

 {\bf Preprint SB/F/258} \vskip .5cm \hrule
\vskip 4.cm
\begin{center}
{\Large{\bf{On the Interacting Chiral Gauge Field Theory in $D=6$
and the Off-Shell Equivalence of Dual Born-Infeld-Like Actions}}}
\end{center}

\vskip.5cm \centerline{{\large A. De Castro and A. Restuccia}}
\vskip .5cm \centerline{\it Universidad Sim\'{o}n Bol\'{\i}var,
Departamento de F\'{\i}sica} \vskip .5cm \centerline{\it Apartado
postal 89000, Caracas 1080-A, Venezuela.} \centerline{\it e-mail:
alex@fis.usb.ve, arestu@usb.ve}

\vskip 2cm
\begin{center}
{\bf Abstract}
\end{center}
\begin{quotation}
\noindent{ A canonical action describing the interaction of chiral
gauge fields in $D=6$ Minkowski space-time is constructed. In a
particular partial gauge fixing it reduces to the action found by
Perry and Schwarz. The additional gauge symmetries are used to
show the off-shell equivalence of the dimensional reduction to
$D=5$ Minkowski space-time of the chiral gauge field canonical
action and the Born-Infeld canonical action describing an
interacting $D=5$ Abelian vector field. Its extension to improve
the on-shell equivalence arguments of dual D-brane actions to
off-shell ones is discussed.}
\end{quotation}

\newpage

\section{Introduction}
A manifestly covariant action for the bosonic $D=11$ five-brane
was constructed in {\cite{re:PST97}} and generalized to a
\textit{kappa} -invariant action for the M-theory super five-brane
in $D=11$ supergravity backgrounds {\cite{re:BLNPST97}}, the
latest theory was independently obtained in {\cite{re:APPS97}}
from a non-manifestly covariant action. The above results are
based on previous work {\cite{re:PS97}}.

The authors of reference {\cite{re:PS97}} constructed a $D=6$
Lorentz invariant interacting theory of a self-dual tensor gauge
field, which gives Born-Infeld theory upon reduction to $D=5$. The
fully covariant action for the bosonic $5$-branes constructed in
{\cite{re:PST97}} may be obtained as an extension of the non
covariant action of reference {\cite{re:PS97}} via the PST
auxiliary scalar field approach {\cite{re:PST95}}
{\cite{re:PST2-1997}}\textit{. } The five-brane superfield
equations were also obtained using the superembedding of the world
volume superspace into a target superspace in \cite{re:HSW97}.

The duality between M-Theory and $IIA$ superstring suggests that
the double dimensional reduction of the $M5$-brane action should
coincide with the duality transformed Dirichlet $4$-brane action,
this was proven in {\cite {re:APPS}}. The approach followed there
is similar to that used in the analysis of duality transformation
in {\cite{re:S96}}{\cite{re:T96}}{\cite {re:AS96}}{\cite{re:L}}.

The essential steps involved in world-volume duality transformation of $D$%
-brane actions may be described for the Born-Infeld theory. The
Born-Infeld action in $n$ dimensional Minkowski space-time is
given by

\begin{equation}
S_A=-\int d^{n}x{\hskip 0.2cm}\sqrt{-det(\eta _{\mu \nu }+F_{\mu
\nu })} \label{eq:1accion}
\end{equation}

\noindent where $F$ is the curvature of the connection $1-$form
$A$. The idea is to formulate the theory in terms of a dual
$(n-3)$-form potential $B$ defined by

\begin{equation}
{}^{*}H=-\frac{\delta S}{\delta F}
\end{equation}

\noindent where ${}^{*}H$ is the Hodge dual of the ``curvature''
$(n-2)$-form

\begin{equation}
H=dB.
\end{equation}

In this formulation the Bianchi identity for the $B$ field is the
field equation for the connection $A$, while the Bianchi identity
for $A$ is the field equation for the potential $B$. The next step
in the approach is to solve for $F$ in terms of ${}^{*}H$ and then
to find the action which gives the field equations for $B$, a
problem which becomes increasingly non-linear and difficult to
solve with the dimensionality. Nevertheless the programme
was successfully carried out in {\cite{re:PS97} and \cite{re:APPS}} for $%
n\leq 5$ and the action for the $B$ field turned out to be

\begin{equation}
S_B=-\int d^nx{\hskip
0.2cm}\sqrt[]{-det(\eta_{\mu\nu}+i{}^*H_{\mu\nu})}.
\label{eq:2accion}
\end{equation}

The equivalence between the action (\ref{eq:1accion}) and
(\ref{eq:2accion}) has been then established over the field
equations. The approach just described was extended in
{\cite{re:APPS}} to prove on-shell equivalence of duality
transformed actions for super D-branes including Wess-Zumino
terms.

In this paper we show the \textbf{off-shell} equivalence of the
canonical actions associated to(\ref{eq:1accion}) and
(\ref{eq:2accion}), which will be called the Born-Infeld action
for a 1-form gauge field and a 2-form gauge field respectively. We
consider the particular case $n=5$ which corresponds to the double
dimensional reduction of the $5$-brane action and the Dirichlet
four-brane action. We start from the action proposed in
{\cite{re:PS97}} for the interacting theory of self-dual tensor
gauge field and look for an action with first class constraints
only. The canonical action we propose in a particular partial
gauge fixing reduces to the action in {\cite{re:PS97}}. The action
is described with the same canonical variables as the one in
{\cite{re:PS97}}. It does not require the introduction of
auxiliary fields. It has more gauge symmetries, but the manifest
$SO(5,1)$ invariance, which was reduced to a manifest $SO(4,1)$ in
\cite{re:PS97} is now reduced to $SO(4)$. We will then consider
the reduction of the action to $5$-dimensional Minkowski
space-time. In a particular admissible partial gauge fixing we
will obtain the canonical formulation of (\ref{eq:1accion}) and in
a different admissible partial gauge fixing we will obtain the
canonical formulation of (\ref{eq:2accion}). The equivalence
between the canonical formulations will then be established
off-shell, without using the field equations. The equivalence
between duality transformed quadratic actions is always an
off-shell equivalence. The difficulties arises when considering
interacting theories, in particular when the actions involved
include higher order time derivative terms as in the case we will
discuss. In \cite{re:L} off-shell arguments between dual actions
have also been used, by realizing the duality symmetry as
canonical transformations. In {\cite{re:Town96}} the standard
duality transformation was used to show the off-shell equivalence
of the $D=11$ M-2-Supermambrane and the $D=10$ IIA D-brane.

We will start our analysis using the non-manifestly covariant
action in {\cite{re:PS97}}, we think the approach will follow also
from the covariant action in {\cite{re:PST97}}. In fact, the
canonical analysis of the covariant action for a free chiral
tensor gauge field in \cite{re:PST2-1997} shows that in this
formulation there are first class constraints only. However, the
constraints as presented in \cite{re:PST2-1997} have a peculiar
behaviour over the submanifold $\partial _{\alpha }a=0$, as
explained in that paper. Since in any case we will break the
manifest Lorentz invariance in our argument, we prefer to start
directly from the action in {\cite{re:PS97}}. Recently, a
canonical analysis of the five-brane action was performed in
{\cite{re:BST98}} and a canonical Hamiltonian obtained. The second
class constraints are, however, still present in that formulation.
The mechanism we will implement to eliminate the second class
constraints of the action in {\cite{re:PS97}}, works exactly in
the same way for the elimination of the second class constraints
of the five-brane actions in {\cite{re:APPS97}}. It seems
important to have a canonical formulation of the five-brane action
with first class constraints only, in order to have control over
all the gauge symmetries and to determine the physical Hamiltonian
of the theory. It is well known that the physical Hamiltonian of
the $D=11$ M-2-supermambrane has very peculiar properties which
yield a continuous spectrum from $0$ to $\infty $ when the
M-2-brane maps the world-volume onto $D=11$ Minkowski target space
{\cite {re:deWLN89}}. The situation may change however when the
membrane maps over compactified target space. Particularly
interesting is the case when the target space is
$M_{9}XS^{1}XS^{1}$. In that case, the infinite dimensional
``valleys'', responsible of the continuity of the spectrum,
disappear at least from the minima of the Hamiltonian
{\cite{re:MRT98}}. In spite of several improvements in the
analysis, the spectrum of the M-2-brane in the compactified case
is still unknown{\cite{re:deWPP97}}. It is then of interest the
analysis of the five-brane physical Hamiltonian, together with a
canonical formulation free of second class constraints. In this
paper we construct that action for the case of the interacting
theory of chiral gauge bosons on 6-dimensional Minkowski
space-time and use its additional gauge symmetries to show the
off-shell equivalence between its reduction to 5-dimensional
Minkowski space-time and the Born-Infeld theory over that base
space.

\section{Analysis of the Free Theory}

We consider a 6-dimensional Minkowski space-time, with coordinates
$x^M$, $M=0,...,5$, where $x^5$ denotes the temporal coordinate
while $x^\mu$, $\mu=0,1,2,3,4$, denote spatial coordinates. The
``curvature" 3-form $H$ in terms of the ``connection" 2-form $B$
is given by:

\begin{equation}
H=dB,
\end{equation}

\noindent{}and in component notation:

\begin{equation}
H=H_{LMN} dx^L\wedge dx^M\wedge dx^N=\frac{1}{3}(\partial_L
B_{MN}+\partial_N B_{LM}+\partial_M B_{NL}) dx^L\wedge dx^M\wedge
dx^N, \label{eq:Hodge6}
\end{equation}

The Hodge dual 3-form $^* H$ has the expression:

\begin{equation}
^* H=\frac{{}\sqrt[]{-G}}{6}\epsilon_{RSTMNP}H^{MNP}dx^R\wedge
dx^S\wedge dx^T
\end{equation}

Where $G$ is in general the determinant of the six dimensional
metric. In our case $-1$.

The self-duality field equations:

\begin{equation}
^* H=H, \label{eq:selfdualcond}
\end{equation}

\noindent were obtained in {\cite{re:PS97}} from an action, which
is not manifestly covariant, expressed in terms of the
5-dimensional Hodge dual to $H$, with components:

\begin{equation}
^*H_{\mu\nu}=\frac{1}{2}{\hskip
0.2cm}\sqrt[]{g}\epsilon_{\mu\nu\rho\lambda\sigma}H^{\rho\lambda\sigma}.
\label{eq:Hodge5}
\end{equation}

In (\ref{eq:Hodge5}) $H^{\rho\lambda\sigma}$ is obtained from the
primitive spatial components $H_{\mu\nu\lambda}$ by raising the
indices with the 5-dimensional spatial components of the metric,
in our case $\delta_{\mu\nu}$. In distinction, the indices in
(\ref{eq:Hodge6}) are raised using the six dimensional metric
$\eta_{MN}$.

The action proposed in {\cite{re:PS97}} is:

\begin{equation}
S_{PS}=\int_{M_6}(^* H^{\mu\nu}\partial_5 B_{\mu\nu}- ^*
H^{\mu\nu}{}^*H_{\mu\nu})d^6x, \label{eq:accionPS}
\end{equation}

\noindent where again the 5-dimensional metric has been used to
raise the indices.

The canonical analysis of (\ref{eq:accionPS}) may be performed in
an straightforward way. The canonical action is constrained by the
mixture of first and second class constraints:

\begin{equation}
\varphi^{\mu\nu}\equiv P^{\mu\nu}-2^* H^{\mu\nu}=0.
\label{eq:2class}
\end{equation}

The canonical Hamiltonian is:

\begin{equation}
{\cal{H}}={\cal{H}}_0+\lambda_{\mu\nu}\varphi^{\mu\nu}
\label{eq:freehamil}
\end{equation}

\noindent where ${\cal{H}}_0=^* H^{\mu\nu}{}^*H_{\mu\nu}$ is
positive definite.

Although the action (\ref{eq:accionPS}) proposed in
{\cite{re:PS97}} yields the required field equations
(\ref{eq:selfdualcond}), it has second class constraints,
therefore it should exist an action with first class constraints
only, from which (\ref{eq:accionPS}) could be obtained by partial
gauge fixing. When we consider dynamical systems with a finite
number of degrees of freedom this is always possible, however,
when the dynamics involves infinite degrees of freedom the new
action may involve non-localities or an infinite number of first
class constraints {\cite{re:GRS91}}. In the case under
consideration, there is however a closed solution to the new
action. It is given, in canonical form, by:

\begin{equation}
S=\int_{M_6}[\frac{1}{2}P^{\mu\nu}\partial_5
B_{\mu\nu}-\frac{1}{4}
({}^*H^{ij}+\frac{1}{2}P^{ij})({}^*H_{ij}+\frac{1}{2}P_{ij})-2^*
H^{0i}{}^*H_{0i}]d^6x, \label{eq:accioncan}
\end{equation}

\noindent{} subject to the first class constraints:

\begin{eqnarray}
\phi^{i}&\equiv & P^{0i}-2 {}^*H^{0i}=0, \label{eq:1class1}\\
 \Lambda^\nu &\equiv&\partial_\mu P^{\mu\nu}=0, \label{eq:1class2}
\end{eqnarray}

\noindent where $i=1,2,3,4$ and $\mu=0,1,2,3,4$.

The first class constraints conmute between themselves and with:

\begin{equation}
{\cal{H}}_0=\frac{1}{4}
({}^*H^{ij}+\frac{1}{2}P^{ij})({}^*H_{ij}+\frac{1}{2}P_{ij})-2{}^*
H^{0i}{}^*H_{0i}, \label{eq:ourhamil}
\end{equation}

We emphasize that $x^0$ is an spatial coordinate in this
formulation.

The constraints (\ref{eq:1class1}) and (\ref{eq:1class2}) are not
independent. In fact

\begin{eqnarray}
\partial_i \phi^i-\Lambda^0&=&0,\label{eq:indep1class}\\
\partial_\mu\Lambda^\mu&=&0\label{eq:indep1class2}
\end{eqnarray}

There are seven independent first class constraints. The original
set of constraints (\ref{eq:2class}) consists of ten independent
constraints. Six of them are second class constraints which come
from the partial gauge fixing of (\ref{eq:1class1}). The whole set
(\ref{eq:2class}) arises then from partial gauge fixing of
(\ref{eq:1class1}), (\ref{eq:1class2}). Moreover in this
particular gauge (\ref{eq:accioncan}) reduces
to(\ref{eq:accionPS}). We have obtained then an action whose
dynamics is restricted by first class constraints only which in a
particular gauge fixing yields (\ref{eq:accionPS}) The new action
has more gauge symmetries, and we will exploit this fact shortly,
however we have lost the manifest Lorentz invariance of
(\ref{eq:accionPS}) on the 5-dimensional space-time. The full
Lorentz invariance will, of course, be recovered at the S-matrix
level.

\section{Equivalence Between the Chiral Gauge Field Theory and Born-Infeld Theory.}
\subsection{The Free Case.}
\label{susec:freecase}

We will discuss in this section the quantum equivalence between
the free theory of a 2-form gauge field and the Maxwell theory
over a 5-dimensional Minkowski space-time. The equivalence arises
directly by using the standard duality procedure which we briefly
review. However, this approach fails as an off-shell argument when
we consider the equivalence of the dimensional reduced interacting
chiral gauge field theory and the Born-Infeld action for a
curvature 2-form $F(A)$. In fact, the equivalence between both
theories has only been established on-shell {\cite{re:PS97}} and
{\cite{re:APPS}}. We will follow a different approach, which may
be extended to show the off-shell equivalence of the interacting
theories under consideration. In this section we apply the
approach to the free theories, to show how it works in a
straightforward case. The standard duality approach, when applied
to the action:

\begin{equation}
S_c=-\int_{M_5}{}^*H^{ab}{}^*H_{ab} d^5x,\hskip .5cm
a,b=1,2,3,4,5,\label{eq:action4db}
\end{equation}

\noindent yields the master action:

\begin{equation}
S_d=\int_{M_5}(-L^{ab}L_{ab}+i F_{ab}(A)L^{ab})d^5x,
\label{eq:acciond}
\end{equation}

\noindent where $F_{ab}(A)$ is the curvature of the connection
1-form $A$. Functional integration on $A$ in (\ref{eq:acciond})
yields

\begin{equation}
\partial_a L^{ab}=0,
\end{equation}

\noindent which implies

\begin{equation}
L^{ab}={}^*H^{ab}.
\end{equation}

After substitution in (\ref{eq:acciond}), we obtain
(\ref{eq:action4db}). If instead we functionally integrate on $L$
in (\ref{eq:acciond}), we obtain the quantum equivalent action:

\begin{equation}
S_d=-\int\frac{1}{4} F_{ab}(A)F^{ab}(A) d^5x. \label{eq:maxaccion}
\end{equation}

(\ref{eq:action4db}) and (\ref{eq:maxaccion}) are then off-shell
equivalent actions. The duality approach as described is very
useful when analyzing global properties of the theories, but has
the drawback that becomes, in most of the cases, useless as an
off-shell argument, when the theories describe non-linear
interactions. We consider now a different argument. We start from
the dimensional reduction of (\ref{eq:accioncan})to 5-dimensional
Minkowski space-time. We compactified $X^0$ and take the zero mode
of the Fourier expansion. We will consider two different partial
gauge fixing conditions. (\ref{eq:accioncan}), (\ref{eq:1class1}),
(\ref{eq:1class2}) under one of the gauge fixing will reduce to
(\ref{eq:action4db}) while under the other one becomes
(\ref{eq:maxaccion}).

We impose the partial gauge fixing condition:

\begin{equation}
B_{0i}=0, \label{eq:agaugefix}
\end{equation}

\noindent associated to the first class constraints
(\ref{eq:1class1}). We may then use (\ref{eq:agaugefix}) and
(\ref{eq:1class1}) to perform a canonical reduction of
(\ref{eq:accioncan}). we obtain:

\begin{equation}
S_1=\int_{M_5}[\frac{1}{2}P^{ij}\partial_5
B_{ij}-\frac{1}{16}P^{ij}P_{ij}-2{}^*H^{0i}{}^*H_{0i}+\rho_j\partial_i
P^{ij}]d^5x,\label{eq:accion1}
\end{equation}

\noindent where we have used

\begin{equation}
{}^*H^{ij}=\frac{1}{2}\epsilon^{ij0kl}\partial_0
B_{kl}+\epsilon^{ijk0l}\partial_k B_{0l}=0
\end{equation}

\noindent and we have introduced the remaining first class
constraints into the action by means of a Lagrange multiplier
which may be reinterpreted as:

\begin{equation}
\rho_i=B_{5i}.\label{eq:maltiplier1}
\end{equation}

Functional integration on $P^{ij}$ gives, after some calculations,

\begin{equation}
S_1=-\int_{M_5}{}^*H^{ab}{}^*H_{ab} d^5x. \label{eq:accion4db1}
\end{equation}

We impose now a different admissible gauge fixing condition.
Instead of (\ref{eq:agaugefix}) we consider:

\begin{eqnarray}
P_T^{ij}&=&0 \label{eq:bgaugefix1}\nonumber \\ B_{Lij}&=&0,
\label{eq:bgaugefix2}
\end{eqnarray}

Where the transverse $T$ and longitudinal $L$ parts are defined in
terms of the derivative operator. That is, for an anti-symmetric
tensor, we have

\begin{eqnarray}
V^{ij}&=&V_T^{ij}+\partial^i V_L^j -\partial^jV_L^i\\
\partial_iV_T^{ij}&=&0,
\end{eqnarray}

\noindent (\ref{eq:bgaugefix1}) together with (\ref{eq:1class1})
and the subset of (\ref{eq:1class2})

\begin{equation}
\partial_jP^{ji}+\partial_0P^{0i}=0
\end{equation}

\noindent allow the elimination of the canonical conjugate pair
$(P^{ij},B_{ij})$. The canonical reduced action is then:

\begin{equation}
S_2=\int_{M_5}[P^{0i}\partial_5
B_{0i}-\frac{1}{4}{}^*H^{ij}{}^*H_{ij}-\frac{1}{2}P^{0i}P_{0i}+\lambda\partial_i
P^{0i}]d^5x,\label{eq:accion2}
\end{equation}

Functional integration of $P^{0i}$ yields, after some
calculations,

\begin{equation}
S_2=-\int\frac{1}{4} F_{ab}(A)F^{ab}(A) d^5x.
\label{eq:max2accion}
\end{equation}

Where $F_{ab}=\partial_aA_b-\partial_bA_a$ and $A_a=B_{0a}$.

We have thus obtained the off-shell equivalence of
(\ref{eq:accion4db1}) and (\ref{eq:max2accion}) since both arise
from different admissible gauge fixing of the action
(\ref{eq:accioncan}), more precisely from its dimensional
reduction to $M_5$ . The factors in front of the actions
(\ref{eq:accion4db1}) and (\ref{eq:max2accion}) are exactly the
same as the ones obtained from the duality transformation.

\subsection{The Interacting Theories}

Following {\cite{re:PS97}} we consider the interacting chiral
gauge field theory governed by the action:

\begin{equation}
S=\int_{M_6}[{}^*H^{\mu\nu}\partial_5 B_{\mu\nu}-4 {\hskip
0.3cm}\sqrt[]{det(\delta_{\mu\nu}+{}^*H_{\mu\nu})}]d^6x.
\label{eq:tbiaccion}
\end{equation}

Since in our formulation $X^5$ is the time coordinate, the metric
to be used inside the square root is $\delta_{\mu\nu}$.
Consequently there is no $i$ factor in front of ${}^*H_{\mu\nu}$,
otherwise an indefinite expression will arise from the terms $1+$
quadratic terms. The $-$ sign in front of the square root is
required to have a positive definite Hamiltonian of the resulting
theory. It is straightforward to see that the expansion of
(\ref{eq:tbiaccion}) up to quadratic terms on the fields agrees
exactly with (\ref{eq:accionPS}). The changes in
(\ref{eq:tbiaccion}) with respect to the action proposed in
{\cite{re:PS97}} are then the necessary ones when $X^5$ is
considered as a temporal coordinate instead of a spatial
coordinate as taken in {\cite{re:PS97}}. The canonical analysis of
(\ref{eq:tbiaccion}) yields the mixture of first and second class
constraints (\ref{eq:2class}) as in the free case. The subset of
first class constraints is (\ref{eq:1class2}), representing four
independent constraints. The remaining six constraints in
(\ref{eq:2class}) are second class ones. The canonical Hamiltonian
is given by (\ref{eq:freehamil}) where ${\cal{H}}_0$ is in this
case

\begin{equation}
{\cal{H}}_0=4{\hskip
0.2cm}\sqrt[]{det(\delta_{\mu\nu}+{}^*H_{\mu\nu})},
\end{equation}

\noindent $\Lambda^\mu$ commutes with ${}^*H^{\sigma\lambda}$ and
hence with ${\cal{H}}_0$. We may now look for an action with first
class constraints only, from which (\ref{eq:tbiaccion}) may be
obtained by partial gauge fixing. It is given by

\begin{equation}
S=\int_{M_6}[\frac{1}{2}P^{\mu\nu}\partial_5 B_{\mu\nu}-4{\hskip
0.2cm}\sqrt[]{det(\delta_{\mu\nu}+M_{\mu\nu})} +
\lambda_i\phi^i+\rho_\nu\Lambda^\nu]d^6x \label{eq:ourtbiaccion}
\end{equation}

Where

\begin{eqnarray}
M_{0i}&=&{}^*H_{0i}\nonumber \\
M_{ij}&=&\frac{1}{2}({}^*H_{ij}+\frac{1}{2}P_{ij}).
\end{eqnarray}

\noindent $\phi^i$ and $\Lambda^\nu$ are given by
(\ref{eq:1class1}) and (\ref{eq:1class2}) respectively. $\phi^i$
and $\Lambda^\mu$ commute between themselves and with the matrix
elements $M_{\mu\nu}$. They hence conmute with the canonical
Hamiltonian.

Again, as in the free case, by gauge fixing of the symmetries
generated by $\phi^i$ and use of the first class constraints, we
obtain

\begin{equation}
P^{\mu\nu}=2{}^*H^{\mu\nu}
\end{equation}

\noindent which after its substitution in (\ref{eq:ourtbiaccion})
yields exactly (\ref{eq:tbiaccion}).

 We will study now the quantum equivalence between the dimensional reduction of the
 interacting theory of chiral
 gauge fields (\ref{eq:ourtbiaccion}) and the Born-Infeld action in $D=5$  Minkowski space-time.

We proceed as in the free case, we will show that there are two
admissible partial gauge fixing of (\ref{eq:ourtbiaccion}) which
give rise in one case to the canonical formulation of the
Born-Infeld action for interacting 1-form gauge fields, and in the
other case to the canonical formulation of the Born-Infeld type
action for interacting 2-form gauge fields. The canonical actions
are equivalent at the quantum level since they will arise from
different admissible gauge fixing of the master action
(\ref{eq:ourtbiaccion}) (provided there is no gauge anomaly in the
quantum formulation of the theory). The off-shell equivalence is
established between the canonical formulations of the actions not
between the covariant Lagrangian formulations. The on-shell
equivalence between the covariant formulations was proven in
{\cite{re:APPS}} and on this aspect we do not have anything to
add.

However, from the point of view of the quantum theory formulated
in terms of functional integrals, one must start with the
canonical formulation for the actions and our equivalence argument
implies then the first step towards the quantum equivalence of the
theories.

We consider first the partial gauge fixing (\ref{eq:agaugefix}),
that is

\begin{equation}
B_{0i}=0
\end{equation}

Its Poisson bracket with $\phi^j$ is

\begin{equation}
\{B_{0i}(x),\phi^j(x^{\prime})\}=\delta^6(x-x^{\prime})\delta^j_i
\end{equation}

We then eliminate $P^{0i}$ from (\ref{eq:1class1}),

\begin{equation}
P^{0i}=2{}^*H^{0i}.
\end{equation}

The canonical reduction of (\ref{eq:ourtbiaccion}) is given by

\begin{equation}
S=\int_{M_6}[\frac{1}{2}P^{ij}\partial_5 B_{ij}-4{\hskip 0.2cm
}\sqrt[]{det(\delta_{\mu\nu}+M_{\mu\nu})}+ \rho_j\Lambda^j]d^6x
\label{eq:ourtbiaaccion}
\end{equation}

\noindent where
${}^*H_{ij}=\frac{1}{2}\epsilon_{ij0kl}\partial^0B^{kl}$.

If we consider the dimensional reduction to 5-dimensional
Minkowski space-time, we then have

\begin{equation}
{}^*H_{ij}=0,
\end{equation}

\noindent and

\begin{equation}
\Lambda^j=\partial_0(2{}^*H^{oj})+\partial_iP^{ij}=\partial_iP^{ij}.
\end{equation}

We notice that, because of (\ref{eq:indep1class}), we are left
with only the four first class constraints

\begin{equation}
\Lambda^j=0
\end{equation}

If we take the quadratic approximation in
(\ref{eq:ourtbiaaccion}), after the reduction to $M_5$ we exactly
obtain (\ref{eq:accion1}). That is, (\ref{eq:ourtbiaaccion})
describes the dynamics of the same number of degrees of freedom of
the interacting 2-form gauge field theory over $M_5$. The Lagrange
multiplier in (\ref{eq:ourtbiaaccion}) may be interprete as
$B_{5i}$ , see (\ref{eq:maltiplier1}). We will now show that it is
in fact, exactly the same dynamics. Moreover, we will now prove
that the reduction to $M_5$ of (\ref{eq:ourtbiaaccion}) is the
canonical formulation of the covariant action
(\ref{eq:2accion}).From (\ref{eq:ourtbiaaccion})we obtain

\begin{equation}
\frac{1}{2}(\partial_5B_{ij}+\partial_jB_{5i}-\partial_iB_{5j}){\hskip
0.2cm}\sqrt[]{det(\delta_{\mu\nu}+M_{\mu\nu})}=\frac{\delta
det(\delta_{\mu\nu}+M_{\mu\nu})}{\delta P^{ij}}
\label{eq:ecuacion1}
\end{equation}

We would like to obtain $P^{ij}$ in terms of $B_{kl}$ from this
equation. The problem is not straightforward since it is a higher
order equation on the $P^{ij}$. However a general solution may be
given. In fact, consider the action

\begin{equation}
-4\int_{M_5}{}\sqrt[]{-det(\eta_{ab}+i{}^*H_{ab})}, \hskip .5cm
a,b=1,2,3,4,5.\label{eq:tbiaccion4db}
\end{equation}

Where, as in \ref{susec:freecase}, the indices $a,b$ are related
to 5-dimensional Minkowski space-time. It exactly reduces to
(\ref{eq:accion4db1}) when the quadratic approximation of the
square root of the determinant is taken. Consider the conjugate
momenta to $B_{ij}$

\begin{equation}
{\cal{P}}^{ij}=6{\hskip
0.2cm}\sqrt[]{-det(\eta_{ab}+{}^*H_{ab})}{\hskip
0.2cm}\frac{\delta det(\eta_{ab}+{}^*H_{ab})}{\delta
\partial_5 B_{ij}}.
\label{eq:calP1}
\end{equation}

The right hand side of (\ref{eq:calP1}) contains higher order time
derivatives of the $B_{ij}$ field.

We now take

\begin{equation}
P^{ij}={\cal{P}}^{ij} \label{eq:p=p}
\end{equation}

\noindent and then, one may check that (\ref{eq:ecuacion1}) is
identically satisfied. This means that (\ref{eq:p=p}) is a
solution of (\ref{eq:ecuacion1}). It is not the unique solution to
(\ref{eq:ecuacion1}), but it seems to be the only one giving rise
to covariant field equations. We now eliminate $P^{ij}$ in terms
of $B_{ij}$ an $\partial_5B_{ij}$ in the reduction to $M_5$ of
(\ref{eq:ourtbiaaccion}). It turns out that

\begin{equation}
det(\eta_{\mu\nu}+M_{\mu\nu})=\frac{[1+(H_{i5})^2-\frac{1}{4!}\epsilon^{ijkl}\epsilon^{mnpq}H_{mi}H_{nj}H_{pk}H_{ql}]^2}{-det(\eta_{ab}+i{}^*H_{ab})}.
\label{eq:subs1}
\end{equation}

The kinetic terms in (\ref{eq:ourtbiaaccion})have the same
denominator as the square root of the determinant, that is

\begin{equation}
[-det(\eta_{ab}+i{}^*H_{ab})]^{\frac{1}{2}}.
\end{equation}

We choose the square root of (\ref{eq:subs1}) as

\begin{equation}
[det(\eta_{\mu\nu}+M_{\mu\nu})]^{\frac{1}{2}}=\frac{[1+(H_{i5})^2-\frac{1}{4!}\epsilon^{ijkl}\epsilon^{mnpq}H_{mi}H_{nj}H_{pk}H_{ql}]}{[-det(\eta_{ab}+i{}^*H_{ab})]^{\frac{1}{2}}}.
\label{eq:sqrt1}
\end{equation}

It combines with the kinetic terms of (\ref{eq:ourtbiaaccion}) to
give in the numerator

\begin{equation}
-4[-det(\eta_{ab}+i{}^*H_{ab})]
\end{equation}

We end up exactly with (\ref{eq:tbiaccion4db}). The field
equations of the action

\begin{eqnarray}
S_5=\int_{M_5}[\frac{1}{2}P^{ij}\partial_5 B_{ij}&-&4{\hskip
0.2cm}\sqrt[]{det(\delta_{\mu\nu}+N_{\mu\nu})}+
\rho_j\Lambda^j]d^5x \label{eq:ourtbiaaccion2}\nonumber \\
N_{0i}&=&{}^*H_{0i}\nonumber\\ N_{ij}&=&\frac{1}{4}P_{ij},
\end{eqnarray}

\noindent and the field equations of (\ref{eq:tbiaccion4db}) are
then equivalent. (\ref{eq:ourtbiaaccion2}) is then a canonical
formulation of (\ref{eq:tbiaccion4db}). The latest contains higher
order time derivatives. Generically the canonical formulation is
in that cases not necessarily unique {\cite{re:LlR95}}.

We now consider the partial gauge fixing (\ref{eq:bgaugefix1}),
that is

\begin{eqnarray}
P_T^{ij}&=&0 \nonumber \\ B_{Lij}&=&0
\end{eqnarray}

\noindent (\ref{eq:bgaugefix1}) is an admissible partial gauge
associated to the constraints (\ref{eq:1class1}),

\begin{equation}
\phi^i=P^{0i}-2{}^*H^{0i}=P^{0i}-\epsilon^{0ijkl}\partial_jB_{kl}=0
\end{equation}

In fact, only the transverse part of $B_{kl}$ appears in that
constraint. (\ref{eq:bgaugefix1}) and (\ref{eq:1class1}) allow to
eliminate the conjugate pair $(B_{Tij}, P_T^{ij})$ from the
canonical formulation. Consider now the constraints

\begin{equation}
\Lambda^i=\partial_0 P^{0i}+\partial_jP^{ji}=0. \label{eq:1class3}
\end{equation}

They only involve the longitudinal part of $P^{ij}$. An admissible
gauge fixing condition associated to it is then
(\ref{eq:bgaugefix2}). (\ref{eq:bgaugefix1}) and
(\ref{eq:1class3}) allow then to eliminate the canonical conjugate
pair $(B_{Lij}, P_L^{ij})$. We notice that , after the reduction
to 5-dimensional Minkowski space-time, (\ref{eq:bgaugefix1}) and
(\ref{eq:1class3}) imply over a contractible manifold. (it is the
case for $M_5$), with appropriate boundary conditions,

\begin{equation}
P^{ij}=0
\end{equation}

The canonical reduction of (\ref{eq:ourtbiaccion}), over $M_5$,
becomes then

\begin{equation}
S=\int_{M_5}[P^{0i}\partial_5 B_{0i}-4{\hskip
0.2cm}\sqrt[]{det(\delta_{\mu\nu}+L_{\mu\nu})} +
\lambda\phi^0]d^5x \label{eq:ourtbibaccion}
\end{equation}

Where $L_{0i}=\frac{1}{2}P_{0i}$ and
$L_{ij}=\frac{1}{2}{}^*H_{ij}$. The only constraint left is the
first class Gaussian one

\begin{equation}
\partial_jP^{0j}=0
\end{equation}

The quadratic approximation of (\ref{eq:ourtbibaccion}) yields
exactly (\ref{eq:accion2}). (\ref{eq:ourtbibaccion}) describes
then the same degrees of freedom of the Born-Infeld action for a
gauge field of rank one. We will now show that not only the
degrees of freedom are the same but the dynamics is exactly
equivalent. We consider, as before, the field equations.

Taking variations with respect to $P^{0i}$, we get

\begin{equation}
\frac{1}{2}[\partial_5B_{0i}-\partial_iB_{05}]{\hskip
0.2cm}\sqrt[]{det(\delta_{\mu\nu}+L_{\mu\nu})}=\frac{\delta
det(\delta_{\mu\nu}+L_{\mu\nu})}{\delta P^{0i}}
\label{eq:ecuacion2}
\end{equation}

We now present a solution to this highly non-linear equation on
$P^{0i}$. We start from the covariant action

\begin{equation}
-4\int_{M_5}{}\sqrt[]{-det(\eta_{ab}+\frac{1}{2}F_{ab})}, \hskip
.5cm a,b=1,2,3,4,5\label{eq:biaccion4db}
\end{equation}

\noindent which exactly reduces to (\ref{eq:max2accion}) when the
quadratic approximation for the square root is used. We consider
now the canonical conjugate momenta to $A_a=B_{0a}$,

\begin{equation}
{\cal{P}}^{i}=2{\hskip 0.2cm
}\sqrt[]{-det(\eta_{ab}+\frac{1}{2}F_{ab})}\frac{\delta
det(\eta_{ab}+\frac{1}{2}F_{ab})}{\delta
\partial_5 B_{0i}}
\label{eq:calP2}
\end{equation}

The numerator of the right hand side member of (\ref{eq:calP2}) is
linear in the time derivatives, but the denominator is a
non-linear expression in terms of them. We now take

\begin{equation}
P^{0i}={\cal{P}}^i \label{eq:p=p2}
\end{equation}

\noindent and replace it into (\ref{eq:ecuacion2}). It is
identically satisfied. Having obtained a solution to
(\ref{eq:ecuacion2}), we substitute it into the action.

It turns out that

\begin{equation}
[det(\delta_{\mu\nu}+L_{\mu\nu})]=\frac{[1+\frac{1}{8}f_{ij}^2+\frac{1}{16}\frac{\epsilon^{ijkl}}{4!}\epsilon^{mnpq}f_{mi}f_{nj}f_{pk}f_{ql}]^2
}{-det(\eta_{ab}+\frac{1}{2}F_{ab})}
\end{equation}

The kinetic terms in (\ref{eq:ourtbibaccion}) have the same
denominator as the square root of the determinant, that is

\begin{equation}
(-det(\eta_{ab}+\frac{1}{2}F_{ab}))^{-\frac{1}{2}}
\label{eq:subs2}
\end{equation}

\noindent we choose the square root of (\ref{eq:subs2}) as

\begin{equation}
[det(\delta_{\mu\nu}+L_{\mu\nu})]^{\frac{1}{2}}=\frac{[1+\frac{1}{8}f_{ij}^2+\frac{1}{16}\frac{1}{4!}\epsilon^{ijkl}\epsilon^{mnpq}f_{mi}f_{nj}f_{pk}f_{ql}]^2
}{[-det(\eta_{ab}+\frac{1}{2}F_{ab})]^{\frac{1}{2}}}
\label{eq:sqrt2}
\end{equation}

It combines with the kinetic terms of (\ref{eq:ourtbibaccion}) to
give in the numerator

\begin{equation}
-4[-det(\eta_{ab}+\frac{1}{2}F_{ab})]
\end{equation}

We notice that there is a different sign in the fourth order term
of (\ref{eq:sqrt2}) compared with the one in (\ref{eq:sqrt1}).
This is so because in (\ref{eq:ourtbiaaccion}) there also
contributions of fourth order of the same form coming from the
kinetic terms. The overall fourth order term has the right
coefficient .In (\ref{eq:ourtbibaccion}) instead, the fourth order
terms coming form the square root, have only space like
derivatives and sum in a direct way to the kinetic terms.

We obtain exactly (\ref{eq:biaccion4db}). (\ref{eq:ourtbibaccion})
is then a canonical formulation of (\ref{eq:biaccion4db}).

We have thus obtained canonical formulations of
(\ref{eq:tbiaccion4db}) and (\ref{eq:biaccion4db})given by
(\ref{eq:ourtbiaaccion2}) and (\ref{eq:ourtbibaccion})
respectively, and show they are equivalent off-shell. We have thus
realized the duality transformation in terms of two different
gauge fixing conditions of a master action. Several interesting
aspects of the duality symmetry in the Born-Infeld and chiral
bosons actions have been discussed in the literature
{\cite{re:GGRS97}}.

\section{Conclusions}

We constructed an action describing the interaction of chiral
gauge fields in 6-dimensional Minkowski space-time, which in a
particular partial gauge fixing reduces to the Born-Infeld type
action found in {\cite{re:PS97}}. The dynamics of the new action
is restricted by first class constrains only, in distinction to
the Perry-Schwarz formulation of the theory. The new action is
used to prove the off-shell equivalence of the canonical
formulation of the dimensional reduced interacting chiral theory
and Born-Infeld theory for 1-form gauge fields, on $D=5$
space-time. In particular it provides a canonical formulation for
the interacting 2-form gauge field theory, whose Lagrangian
depends on higher order time derivatives. The action we presented,
over $M_5$, has the peculiar property that there are two
admissible partial gauge fixing which give rise to two covariant
theories. The usual situation is that for a given canonical action
there is only one gauge fixing giving rise to a covariant action.
It is clear that the peculiarity is due to the duality
transformation between the two covariant actions. It suggests then
that the approach we presented could be generalized and used to
improve the on-shell equivalence of non-linear dual theories to
off-shell ones.

From the point of view of the quantum formulation of the fields
theories in terms of the functional integral approach, one has to
start with their canonical actions. The off-shell equivalence
between the canonical formulations yields then a first step
towards the proof of the quantum equivalence between the dual
theories. We notice that, since the dependence on the momenta of
the canonical action is not quadratic, we cannot functionally
integrate the momenta to obtain the functional integral in terms
of the covariant Lagrangian. From this point of view, the
off-shell equivalence between the canonical actions is the one
that has to be established.

The elimination of the second class constraints of the action
presented in {\cite{re:APPS97}} may be performed on the same lines
we have presented. We expect the same programme will then follow
to show the off-shell equivalence of dual D-brane actions.

\end{document}